\documentclass[fleqn,10pt,twoside]{article}

\usepackage[english] {babel}

\usepackage {graphicx}
\usepackage {graphics}
\usepackage {floatflt}
\usepackage {latexsym}
\usepackage {caption3} 
\usepackage[labelsep=period]{caption}[2007/11/04 v.3.1e]

\def\noi{\noindent}

\renewcommand{\thesubsubsection}%
        {\arabic{section}.\arabic{subsection}.\arabic{subsubsection}.}

\newcommand{\heads}[2]{\markboth{\protect\small\it #1}{\protect\small\it #2}}

\newcommand{\Arthead}[5]{ \setcounter{page}{#4}\thispagestyle{empty}\noi
    \unitlength=1pt \begin{picture}(500,40)

        \put(0,58){\shortstack[l]{\small\it Gravitation \& Cosmology,
                        \small\rm Vol. #1 (#2), No. #3, pp. #4--#5    \\
        \footnotesize {Proceedings of the International Conference on Gravitation, Cosmology, Astrophysics and Nonstationary Gas Dynamics,}    \\
\footnotesize {Delicated to Prof. K.P.Staniukovich's 90th birthday, Moscow, 2-6 March 2006}    \\
\footnotesize\copyright \ #2 \ Russian Gravitational Society} }

    \end{picture}
	 }     		
\def\prepno#1#2
    {\thispagestyle{empty}
    \noi    \unitlength=1mm
    	\begin{picture}(178,10)
            \put(177,15){\llap{\bf #1}}
            \put(177,10){\llap{\small\rm #2}}
        \end{picture}
          }             

\newcommand{\Title}[1]{\noi {\uppercase{\Large #1}}     }
\newcommand{\Author}[2]{\noi{\large\bf #1}\\[2ex]\noindent{\it #2}   }

\newcommand{\Abstract}[1]{\vskip 2mm \begin{center}
        \parbox{16.4cm}{\small\noi #1} \end{center}\medskip}

\newcommand{\foom}[1]{\protect\footnotemark[#1]}

\newcommand{\email}[2]{\footnotetext[#1]{e-mail: #2}
		\addtocounter{footnote}{1}}

\heads{A.M.Baranov}
      {Gravitational fields of lightons and helixons in General Relativity}

\begin{document}
\twocolumn 
[
\Arthead{12}{2006}{2-3 ({\bf{46-47}})}{100}{102}

\Title{GRAVITATIONAL FIELDS OF LIGHTONS AND HELIXONS 

\vspace{0.2cm}
IN GENERAL RELATIVITY}

\vspace{.5cm}
   \Author{A.M.Baranov\foom 1 } 
{\it Dep. of Theoretical Physics,Krasnoyarsk State University,
79 Svobodny Av., Krasnoyarsk, 660041, Russia}

{\it Received 10 May 2006}

\Abstract
    {A lightlike limit procedure for massive particles of Schwarzschild, Kerr and NUT which are sources of the exterior gravitational fields of an algebraic type D is introduced. It is shown when a velocity of fast moving massive particles along $z$ axis tends towards the light velocity axis and the total energy of each particle is constant (i.e. a rest mass of the particle tends towards zero) together with Kerr's angular momentum along z axis and with NUT's parameter which also tend to constants then the gravitational fields of these fast moving particles tend towards the wave's fields of the N and III algebraic types as their limits. The lightlike limit of massive particle may be described as cusp catastrophe on the level of Weyl's matrix with a change of gravitational field's symmetry of such source. In considered cases such limit is a phase transition of the gravitational field from D type into N type or III type (transition from one "phase" to another). Petrov's algebraic types are different "phases" of gravitational field. As result of the lightlike limit procedure the lightlike scalar massless particle ({\it lighton}) and a vector lightlike massless particle with a helicity ({\it helixon}) are obtained. It is shown the lightlike sources in General Relativity "have no hairs"}
]

\email 1 {bam@lan.kras.ru}

\section{Introduction}

In classical electrodynamics the problem of finding the field of an electric charge when its velocity tends to the velocity of light $c$ is well known [1]. The limiting 
field of such a rapidly moving charge particle tends to the field of a monochromatic electromagnetic plane wave. A similar problem in general relativity for Petrov's symmetric $6 \times 6$ curvature matrix of the gravitational field of a rapidly moving particle is considered in [2]. 

However a correct solution of these two problems is connected with the use of generalized functions (Dirac's $\delta$ functions) [3]. In this case, the velocity $v$ tends to $c$ (here we have $v \rightarrow c = 1$), and the rest mass tends to zero ($m_0 \rightarrow 0$), so that the total relativistic energy of the particle is a constant, $E \rightarrow const $. We shall call this procedure {\it the lightlike limit}. Weyl's matrix is introduced as Petrov's symmetric $3 \times 3$ traceless complex matrix in 3D Euclidean space. We shall call the similar limit on the level of Weyl's matrix eigenvalues {\it the lightlike limit on the level of Weyl's matrices}. Such lightlike limit may be described as a cusp catastrophe.

\section{Lightlike limit of Weyl's \\
matrices} 

The lightlike limit procedure on the level of Weyl's matrices is applied to exterior gravitational fields of algebraic type D of massive particles, i.e. the Schwarzschild, Kerr and NUT fields. In this procedure the velocity of rapidly moving particles tends to $c$ ($\,v \rightarrow 1\,$) along $z$ axis whlie the total energy $E$ of each particle is constant (i.e. the particle rest mass tends to zero, $m_0 \rightarrow 0,\,E= const$) together with Kerr's angular momentum $L_z$ and with NUT's parameter $b,$ which are also kept constant along $z$ axis ($L_z = a\cdot m_0 \rightarrow J\cdot E$ and $b \rightarrow B = const$) [3-7]. As a result of this procedure we have two sorts of massless lightlike particles: scalar particles ({\it lightons}) and spinning particles with helicity ({\it helixons}).

Weyl's matrix 
$$
\hat{W}_{Sch} = (m_0/r_0^3)
\left(\begin{array}{ccc}
 2 & 0 & 0\\
 0 &-1 & 0\\
 0 & 0 &-1
\end{array}\right)
\eqno{(1)}
$$
of Schwarzschild's exterior gravitational field at rest may be transformed into a moving system of reference as 
$$
\hat{W} = \hat{T}\hat{W}_{Sch} \hat{T}^{-1} =
$$
$$ (E/\varepsilon^2/R^3)(\varepsilon^2\hat{C}_D + 3(\hat{W}_N^{(E)} + iv \hat{W}_N^{(B)}))
\eqno{(2)}
$$

\noindent
with $\hat{C}_D = diag(-1,2,-1);\,\hat{W}_N^{(E)}\,$ is {\it an electric} part 
and $\,\hat{W}_N^{(B)}\,$ is {\it a magnetic} part of Weyl's matrix of the gravitational plane wave [3]
$$
\hat{W}_N = l\tilde{l} =
\left(\begin{array}{ccc}
 0 & 0 & 0\\
 0 & 1 & i\\
 0 & i &-1
\end{array}\right) = \hat{W}_N^{(E)}+i \hat{W}_N^{(B)},
\eqno{(3)}
$$ 

\noindent
where $\, i^2 = -1;\;\,\tilde{l} = (1,i,0)$ is an eigenvector;$\,\tilde{l} l = 0;\,$ $\varepsilon= (1-v^2)^{1/2};$ $\hat{T}=(T_1, T_2, T_3)$ is an orthogonal matrix with 
orthogonal vector-coloumns: 
$$
\tilde{T}_1 = (\cosh{\psi}, i\sinh{\psi}, 0); 
$$
$$
\tilde{T}_2 = (-i\sinh{\psi}, \cosh{\psi}, 0);
\eqno{(4)}
$$
$$
\tilde{T}_3 = (0, 0, 1),
$$

\noindent 
with $\cosh{\psi} =1/\varepsilon;\, \sinh{\psi} =
v/\varepsilon.$

As $\,v \rightarrow 1,\,$ $\,\psi \rightarrow \infty.$
Under the procedure of a {\it lightlike limit} 
$$
\hat{W} \rightarrow (6E/\rho^2)\,\delta (z+t) \, \hat{W}_N,
\eqno{(5)}
$$

\noindent
where $\,\delta(z+t)\,\hbox{is}\,\;\delta$ function; $R^2 = 
\rho^2\varepsilon^2 + (z+vt)^2;\, \rho^2 = x^2 + y^2.$ 
So we have in the limit a gravitational field of a scalar massless lightlike particle [3]. This particle is called a {\it lighton} [4].

We have a change in the symmetry of the solutions to the gravitational equations in such a lightlike limit. The Schwarzschild-like limiting metric as a result of the lightlike limit, may be written as [3]
$$
g_{\mu \nu} = \delta_{\mu \nu} -8H l_{\mu}l_{\nu},
\eqno{(6)}
$$

\noindent
where $\mu, \nu = 0,1,2,3;\;$ $\,\delta_{\mu \nu}=diag(1,-1,-1,-1);\;$ 
$\,H=-2E \delta(z+t) ln(\rho^2);\;$ $\;l_{\mu}=\delta_{\mu}^0 +\delta_{\mu}^3;\;$ $\,l_{\mu}l^{\mu}=0.$

The metric (6) describes a singular source and is an exact solution of Einstein's exact equations [3]
$$
(\displaystyle\frac{\partial^2}{\partial t^2}-\Delta)\,g_{\mu \nu} = 8\pi T_{\mu \nu},
\eqno{(7)}
$$

\noindent
where $(\displaystyle\frac{\partial^2}{\partial t^2}-\Delta)$ is d'Alembert's operator in Minkowski space-time, 
$$
T_{\mu \nu} = 2E \delta(z+t)\delta(x)\delta(y)l_{\mu}l_{\nu}
\eqno{(8)}
$$
 is a singular energy-momentum tensor of null radiation which describes a lightlike singular source.

The metric (6) has two Killing's vectors: the {\it lightlike vector} $\xi_L = 
(\partial/\partial t + \partial/\partial z)$ and the {\it spacelike vector} 
$\xi_Z = (x\partial/\partial y - y\partial/\partial x)$ which defines axial symmetry and is equal to $\partial/\partial \varphi$ in polar coordinates. 

The Schwarschild-like solution has four Killing's vectors: the {\it timelike vector} 
$\xi_T =\partial/\partial t$ and three {\it spacelike vectors}: $\xi_X = (y\partial/\partial z - z\partial/\partial y);\;$ $\xi_Y=(z\partial/\partial x - x\partial/\partial z);\;$ $\xi_Z=
(x\partial/\partial y - y\partial/\partial x).$ Further, the Lorentz boost $\hat L$ is applied to the Killing vectors of Schwarzschild's solution, and in the limit of velocity of light, the vector $\xi_Z$ will be invariable while the vectors 
$\hat L \xi_T,\;\hat L \xi_X,\;\hat L \xi_Y$ degenerate into the null vector $\xi_L.$ 

In the case of the NUT solution, similar transformations and procedure of a lightlike limit reduce to 
$$
\hat{W} = \hat{T}\hat{W}_{NUT} \hat{T}^{-1} \rightarrow 6((E-iB)/\rho^2)\,\delta(z+t) W_{N}\rightarrow 
$$
$$
\rightarrow (6E/\rho^2)\,\delta(z+t) W_{N},
\eqno{(9)}
$$ 

\noindent
where the NUT parameter can be eliminated by a choice of the rotation angle, $\varphi = -Arctan(B/E)$ in the complex plane and by a new choice of the parameter $\sqrt{E^2+B^2} \rightarrow E,$ i.e. we have again the gravitational field of a {\it lighton} [5-7].

Analogous transformations and the procedure of a lightlike limit reduce Weyl's matrix of Kerr's exterior gravitational field of a spinning massive particle to 
$$
\hat{W} = \hat{T}\hat{W}_{Kerr} \hat{T}^{-1} \rightarrow
$$
$$
\rightarrow (3/\rho^2)\,\delta(z+t) ( 2E \hat{C}_{N} +(\pi J/2\rho)\sin{\theta} \hat{C}_{III}),
\eqno{(10)}
$$ 

\noindent
where $\hat{C}_{N}=\tilde{l}_{1}l_{1}$ with the eigenvector $\tilde{l}_1=(1,-i,0),$ 

$$
\hat{C}_{N}=
\left(\begin{array}{ccc}
 1 &-i & 0\\
-i &-1 & 0\\
 0 & 0 & 0
\end{array}\right)
\eqno{(11)},
$$

and 
$$
\hat{C}_{III}=(1/2)(\tilde{l}_{2}l_{2}+ \tilde{l}_{3}l_{3}) =
\left(\begin{array}{ccc}
 0 & 0 &-1\\
 0 & 0 & i\\
-1 & i & 0
\end{array}\right)
\eqno{(12)}
$$

\noindent
is Weyl's matrix of algebraic type III with the eigenvectors $\tilde{l}_{2}= (0,1,i)$ and $\,\tilde{l}_{3}=(0,i,1).$ [4,7,8] 

On the other hand, a superposition of two Weyl's matrices of algebraic types N and III is a resulting matrix of algebraic type III (wave type) [9].

A lightlike limiting metric which corresponds to the limiting matrix (10) [3] has two Killing's vectors only: the {\it null vector} $\xi_L = 
(\partial/\partial t + \partial/\partial z)$ and the {\it axial spacelike vector} 
$\partial/\partial \varphi$ in polar coordinates. The Kerr solution has two Killing vectors: the {\it timelike vector} $\xi_T=\partial/\partial t\;$ and the {\it axial spacelike vector} $\xi_Z=\partial/\partial \varphi.$ 

The Lorentz boost is applied to Killing vectors of Kerr's solution together with the lightlike procedure and leads to $\xi_Z \rightarrow 
\xi_Z$; $\;\xi_T \rightarrow \xi_L.$ 

In the limit, we have two massless lightlike particles. In the first case of Kerr's particle with an angular momentum along the $z$ axis we obtain a spinning massless lightlike particle with helicity $L_z = JE$ [7,9]. Such particle is called a {\it helixon} [4]. 

In the second case, when Kerr's relative angular momentum $a=L_{z}/m_{0}$ is perpendicular to the $z$ axis, then the lightlike limit procedure leads to a loss of limiting value of Kerr's relative angular momentum [7,9] and we obtain a {\it lighton}.

\section{The lightlike limit and \\
catastrophe theory}

From the point of view of catastrophe theory such a {\it lightlike limit}
is a {\it catastrophe}. Let $\,{\bf \hat{W}}X = \lambda X\,$
(an eigenvalue problem), then the characteristic equation is 

\begin{figure}[t,h]
\fbox{\parbox{8.2cm}{\rule[-0.5cm]{0 mm}%
{8.2cm}\hfil\centering\includegraphics[width=8cm]{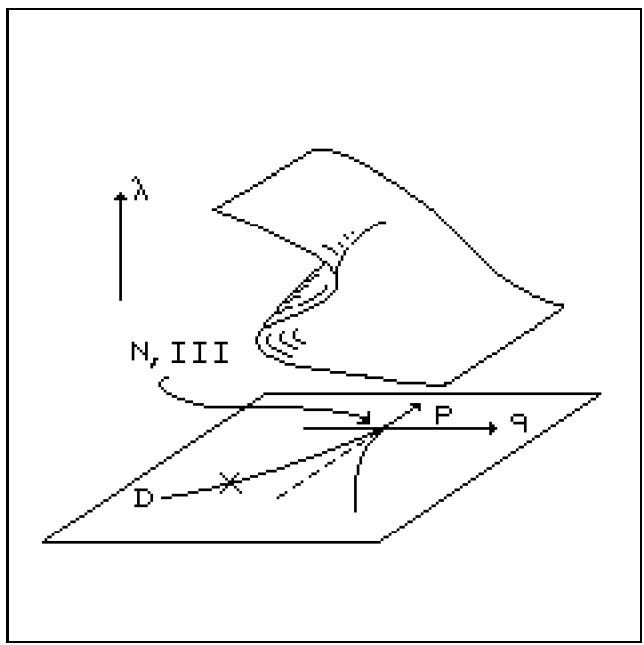}}}
\caption{The cusp catastrophe's surface and its projection onto the plane of control parameters $p$ and $q$.}
\end{figure}

$$
det({\bf \hat{W}}-\lambda {\bf \hat{I}}) \sim\lambda^3 +p\lambda +q = 0\,
\eqno{(13)}
$$

\noindent
with the control parameters $\,p = p(\varepsilon),$ $q = q(\varepsilon)$ and the potential function 
$$
V=(1/4){\lambda}^4+(1/2){\lambda}^2 p+{\lambda}q,
\eqno{(14)}
$$

\noindent
i.e. we have a cusp catastrophe. 

The discriminant of the equation (14) is $\;Q = (p/3)^3 + (q/2)^2. \;\;$ When $Q=0,$ we have the semicubical parabola $p = -3(q/2)^{2/3}$ which corresponds to Weyl's matrix of type D (see Fig.1). Our case is marked by a cross ($q < 0$).

As $v \rightarrow 1,$ we have $\varepsilon \rightarrow 0 $
 and $p(\varepsilon) \rightarrow 0,\; q(\varepsilon ) \rightarrow 0 $. 
From the point of view of second-order phase transitions, a cusp point ($p = q
= 0$) is here the point of a phase transition of the Schwarzschild [10] or NUT 
gravitational fields from type D into type N and Kerr's gravitational
field from type D into types N or III. This is a transition from one "phase" of the gravitational field to another. The parameter $\,p\,$ plays the role of temperature. 
The derivative $\,\partial V/\partial p\,$ plays the role of entropy, and $\,\partial^2 V/\partial p^2\,$ corresponds to thermal capacity.

All physical parameters are lost under such a limiting procedure, except for the total energy $E$ and the helicity $JE$. So we may say that lightlike sources in general relativity "have no hairs" [7].

\section{Summary}

This paper summarizes some results concerning the lightlike limits of the Schwarzschild, Kerr and NUT solutions on the level of Weyl matrices and investigations with catastrophe theory. Under such a lightlike limit procedure, the algebraic type of the gravitational field changes, i.e. the original symmetry of the gravitational field is broken. The lightlike procedure leads to two lightlike particles: the lighton (scalar particle) and the helixon (spinning particle with helicity).

\end{document}